\documentclass[twocolumn,prl,superscriptaddress,showkeys 
]{revtex4-1}

\usepackage[fleqn]{amsmath}
\usepackage{amssymb}
\usepackage{bm}
\usepackage{graphicx} 
\usepackage{subfigure}
\RequirePackage{color}
\begin{document}

\title{Direct and noisy transitions in a model shear flow}

\author{Marina Pausch}
\affiliation{Fachbereich Physik, Philipps-Universit\"at Marburg, 35032 Marburg, Germany}
\author{Bruno Eckhardt}
\affiliation{Fachbereich Physik, Philipps-Universit\"at Marburg, 35032 Marburg, Germany}
 \affiliation{J.M. Burgerscentrum, Delft University of Technology, 2628 CD Delft, The Netherlands}


\begin{abstract}
\noindent \textbf{Abstract} 
The transition to turbulence in flows where the laminar profile is linearly stable requires perturbations 
of  finite amplitude. ``Optimal" perturbations are distinguished as extrema of certain functionals, and
different functionals give different optima. We here discuss the phase space structure of a 2-d
simplified model of the transition to turbulence and discuss optimal perturbations with respect to
three criteria: energy of the initial condition, energy dissipation of the initial condition and
amplitude of noise in a stochastic transition. We find that the states that trigger the transition are
different in the three cases, but show the same scaling with Reynolds number.
\end{abstract}


\maketitle
\section{Introduction}
In parallel shear flows like pipe flow, plane Couette flow or Poiseuille flow and in boundary layers like
the asymptotic suction boundary layer or the Blasius profile, turbulence appears when the laminar
profile is linearly stable against perturbations \cite{Grossmann:2000}. 
Accordingly, finite amplitude perturbations are required to 
trigger turbulence, a scenario referred to as by-pass transition \cite{Morkovin:1969tj}. 
Many studies in the above flows have 
shown that the transition to turbulence is associated with the presence of 3-d exact coherent states
\cite{Eckhardt:2007ix}.
They appear in saddle-node bifurcations which in the state space of the system create regions 
of initial conditions that do not decay to the laminar profile, but instead are attracted
towards the node-state \cite{Kreilos:2012bd}. As the Reynolds number increases, 
the region widens, the node state undergoes further bifurcations and chaotic attractors 
or saddles are formed \cite{Mellibovsky:2011gq,Mellibovsky:2012kf,Halcrow:2009jx}. 
Initial conditions can only trigger turbulence when they reach into that interior region, 
i.e. cross the stable manifold of the saddle state on the boundary of the region \cite{Skufca:2006iu}. 
An ``optimal" perturbation is one that can trigger turbulence and at the same time
is a minimum of a prescribed functional. Popular is an optimization based on amplification or
energy gain over a given interval in time 
\cite{Biau:2009bb,Pringle:2010gp,Pringle:2012cb,Rabin:2011vm,Rabin:2012ib,Cherubini:2010cp,
Cherubini:2010ju,Cherubini:2013,Cherubini:2014wc} or on the total time-averaged dissipation 
\cite{Monokrousos:2011cq,Duguet:2013}.
Because they take the time-evolution into account, they connect to optimization problems
in control theory \cite{Kerswell:2014vk,Luchini:2014fv}. 

We simplify matters here and focus on a geometric optimization by identifying 
initial conditions that will eventually become turbulent, without regard
of the time it takes for them to become turbulent. The states are optimized so that a certain
quadratic function, such as energy content or dissipation, is extremal: it is a maximum in the sense
that all initial conditions with a lower value of the quadratic function will not become turbulent,
and it is minimal in that the first initial conditions that become turbulent have values larger than
this optimum. 
At the optimal value there will then be at least one trajectory which neither becomes turbulent nor
returns to the laminar profile: it lies on the stable manifold of the edge state \cite{Skufca:2006iu}, so that
the optimum is reached when the iso-contours of the optimization functional touch the stable manifold
of the edge state (similar descriptions of the state space structure can be found in 
\cite{DauchotManneville97,cossu2005optimality,Duguet:2013,Cherubini:2014wc}).

\section{The Model}

To fix ideas and to keep the mathematics as simple as possible, we take the 2-d model introduced
by Baggett and Trefethen \cite{Baggett:1997}.  The model we use is one of a set of many 
low-dimensional models of various levels of complexity
\cite{Gebhardt:1994,Waleffe:1995,DauchotManneville97,Eckhardt:1999,Moehlis:2004il,Moehlis:2005gy,Lebovitz:2009fb,Lebovitz:2013,Kerswell:2014vk}.
It has a non-normal linear part and an energy conserving nonlinearity, and, this being the
most important feature for the present application, it is 2-d so that the
entire phase space can be  visualized (a property it shares with the illustrative model of 
\cite{Kerswell:2014vk}). Despite its simplicity, the model can be used 
to illustrate several features of the transition mechanisms in  shear flows.

The model has two variables, which may be thought of as measuring the amplitudes of 
streaks $x$ and vortices $y$ (see also \cite{Eckhardt:2003cy}),
and one parameter $R$ that plays the role of the Reynolds number:
\begin{eqnarray}
\dot x &=& - x/R + y - y \sqrt{x^2+y^2}\\
\dot y &=& -   2y/R +  x \sqrt{x^2+y^2}
\end{eqnarray}
In order to highlight more clearly what happens near the origin, we magnify
by rescaling the variables with the Reynolds number $R$ (see \cite{Eckhardt:1998ua}), 
i.e. we redefine the amplitudes $x=x'/R^2$, $y=y'/R^2$ and the time $t=R t'$ 
such that (with the primes dropped)
\begin{eqnarray}
\dot x &=& - x + R y - y \sqrt{x^2+y^2}/R\\
\dot y &=& -   2y +  x \sqrt{x^2+y^2} /R
\end{eqnarray}
Time evolution under the nonlinear terms alone preserves $x^2+y^2$, which may be
thought of as a kind of energy, so that the nonlinear terms are ``energy" conserving. 
For $R<R_c=\sqrt{8}$ the only fixed point is $x=y=0$, henceforth referred to 
as the ``laminar" fixed point. At $R=R_c$,  symmetry
related fixed points appear at  $(x_c,y_c)$ and $(-x_c,-y_c)$, with
\begin{eqnarray}
x_c &=& R (2R\pm 2 \sqrt{R^2-8})/D_\pm\\
y_c &=& R (R^2-4\pm R \sqrt{R^2-8})/D_\pm 
\end{eqnarray}
where $D_\pm=\sqrt{8+2 R^2 \pm 2R\sqrt{R^2-8}}$. The two fixed points closest to
the origin are unstable, hence are saddle states, and the two further out
are stable and hence nodes. The saddle states are the ``edge states" 
\cite{Skufca:2006iu} and the node states are in the regions where turbulence would
form, if more degrees of freedom where available. Nevertheless, we will
refer to them as the ``turbulent" states.

\begin{figure}
\includegraphics[width=\columnwidth]{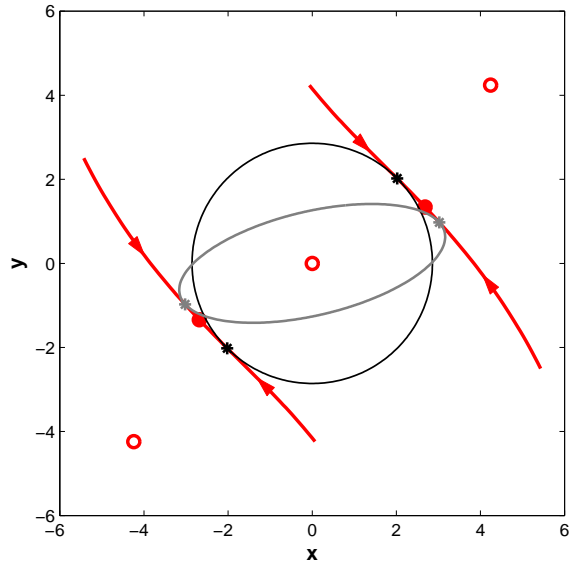}
\caption{\label{fig1} 
State space of the 2-d model for the transition to turbulence for $R=3$. 
The open symbols mark the stable fixed point at the origin (``laminar" state)
and the two nodes from the bifurcation (``turbulent" states). The full symbols
are the edge states, and the red lines indicate the stable manifolds of
the edge states. The black circle and the grey ellipsoid indicate the states
where the energy \eqref{eq:energy} and the noise functional \eqref{eq:Qnoise} are minimal, 
respectively. The points where they touch the stable manifolds are indicated by stars.
}
\end{figure}

As $R\rightarrow \infty$, the saddles are to leading order in $1/R$ located
at $\pm (2, 2/R)$,
which in the original coordinates represents an approach to the origin
like $\pm (2/R^2, 2/R^3)$. The stable manifolds rotate so as to become 
parallel to the $x$-axis, as we will see in the following.

\section{Optimal initial conditions of minimal energy}
The Euclidean distance to the origin can be obtained from a quadratic form
\begin{equation}
E=(1/2)(x^2+y^2)
\label{eq:energy}
\end{equation}
which has the form of a kinetic energy. This assignment is further supported by
the observation that $E$ is preserved under time evolution by the nonlinear terms alone.
In the sense described in the introduction, optimality with respect to this
energy functional thus means the largest value up to which all trajectories
return to the laminar state, and the smallest one where the first trajectories
that evolve towards the turbulent state become possible. On the boundary between 
these two cases are states that neither return to laminar nor become turbulent, 
that lie on the stable manifold of the edge state. Geometrically, we are thus looking
for the circle with largest radius that we can draw around the origin that just touches 
the stable manifold. 
Algorithmically, we find this point by a modified edge tracking which minimizes 
the energy \eqref{eq:energy} as described in the appendix.

An example of such an optimal circle is given in Figure 1, and its variation with $R$ is shown in Figure 2.
As the Reynolds number increases, the fixed point moves towards $(2,0)$ on the abscissa, and the
stable manifold rotates to being parallel to the abscissa. The point of contact between circles of 
equal energy and the stable manifold moves  away from the edge state,
approaches the $y$-axis and moves inwards to the origin like $1/R$. 

\begin{figure}
\includegraphics[width=\columnwidth]{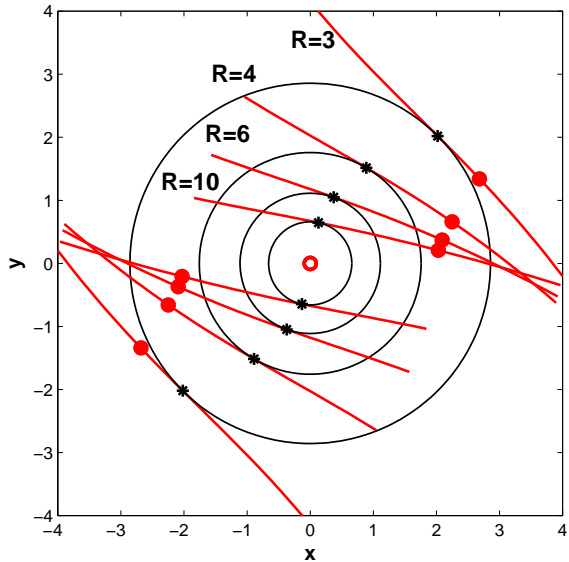}
\caption{\label{fig2}  
Optimal states in energy for different $R$. The open symbol in the middle is the laminar state. States of fixed energy are 
indicated by circles, and the points where they touch the stable manifolds (red lines)
of the edge states (indicated as full symbols) are the points marked by stars. One notes that as 
$R$ increases, the manifolds become more parallel to the 
$x$-axis, and the point of contact approaches the origin from the $y$-axis.
}
\end{figure}

In an insightful discussion of the energy functional, Cossu \cite{cossu2005optimality} notes that
in the time-derivative of the energy functional only the linear parts of the equations of motion
remain and that the nonlinear ones drop out because energy is preserved. This observation allows 
to define a necessary condition for the location of the extremum, which for the two-dimensional example
studied here implies that the optimum lies along the line connecting the laminar and the turbulent 
fixed points. One could then find the optimum by a one-dimensional search along this line. However,
we did not pursue this further, as we also want to find optima with respect to other functionals
that are not preserved by the nonlinear terms.

\section{Optimal initial conditions of minimal energy dissipation}
The diagonal terms in the linear part of the equations of motion correspond to 
the dissipation in the original Navier-Stokes equation. Accordingly, we can define
a dissipation functional \cite{Monokrousos:2011cq}
\begin{equation}
\epsilon=(1/2)(x^2+2y^2)\,.
\label{eq:dissip}
\end{equation}
and study initial conditions that are minimal or optimal with respect to this
functional. As in the previous example, the geometrical condition is that we 
now have to find the point where an ellipse touches the stable manifolds. 
This gives the ellipses shown for different $R$ in
Figure 3. Note that the points where the ellipses touch the stable
manifolds are different from the ones of the energy functional, but their asymptotic
behaviour for large $R$ seems to be similar (see below).

\begin{figure}
\includegraphics[width=\columnwidth]{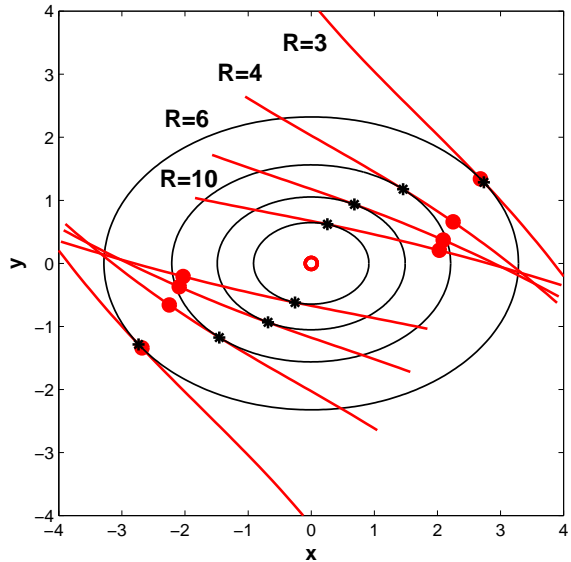}
\caption{\label{fig3} 
Optimal states with respect to the dissipation functional for different $R$. States of fixed dissipation are 
indicated by ellipses, and the points where they touch the stable manifolds (red lines)
of the edge states (full symbols) are marked by stars. The open symbol in the middle 
is the laminar state. One notes that as  $R$ increases, the point of contact moves 
very much as in the case of the energy functional.
}
\end{figure}

\section{Optimal noisy transitions}
As a third example we consider noise-driven transitions. To this end, the equations of motion are
expanded to include a stochastic forcing of the individual terms,
\begin{eqnarray}
\dot x &=& -  x + R y - y \sqrt{x^2+y^2}/R + \xi_x\\
\dot y &=& -   2y + x \sqrt{x^2+y^2}/R + \xi_y
\end{eqnarray}
where the noise is characterized by $\langle \xi_i\rangle=0$ and 
$\langle \xi_i(t) \xi_j(t') \rangle=D_{ij}\delta_{i,j}$. 
We consider the case $D_{11}=D_{22}=D$, so that both components are driven with
equal noise amplitude.
In a linear approximation around the
laminar fixed point, the non-normal coupling between the two components results in 
a probability density function (pdf) for the two components that is Gaussian with a 
covariance matrix given by \cite{Risken:1996} 
\begin{equation}
p(x,y)=\frac{1}{\pi D} [\det(Q)]^{\frac{1}{2}} \exp(-Q_N(x,y)/D)
\label{eq:noise}
\end{equation}
with
\begin{equation}
Q=\frac{1}{2(R^2+9)} 
\begin{pmatrix}
9 & -3 R \\ 
-3 R & 3 R^2+18
\end{pmatrix}
\label{eq:Qnoise}
\end{equation}
and
\begin{eqnarray}
Q_N(x,y)&=& {\bf x}^T Q \, {\bf x} \\
&=&\frac{3}{2(R^2+9)} (3 x^2 - 2 R x y + (R^2+6) y^2).
\label{QN_def}
\end{eqnarray}
Asymptotically, for $R\rightarrow \infty$, the quadratic form becomes 
$Q_N(x,y)\rightarrow 3y^2/2$, so that the Gaussian stretches out
along the $x$-direction for increasing $R$.

\begin{figure}
\includegraphics[width=\columnwidth]{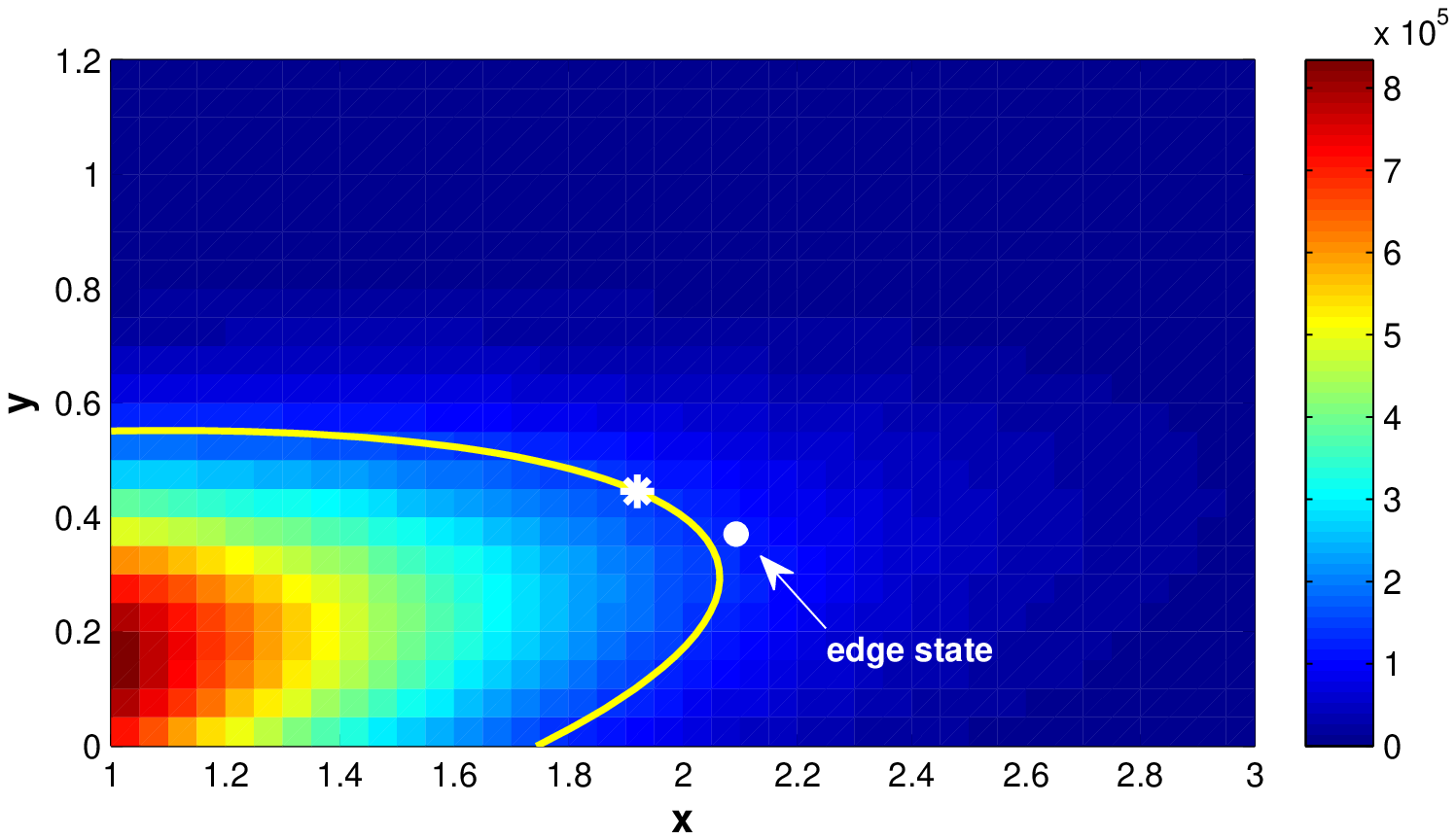}
\caption{\label{fig4} 
Probability density for the linearized equations with noise for $R=6$ in the region of phase space where the transition is expected to occur. For the statistics we calculated the time evolution of 20000 initial conditions starting at the laminar state for 20 time units with a step size of $dt=10^{-3}$. It can be seen that the iso-contours $p=const$ are of elliptical shape. The star indicates the point where the noise functional touches the stable manifold of the edge state.
}
\includegraphics[width=\columnwidth]{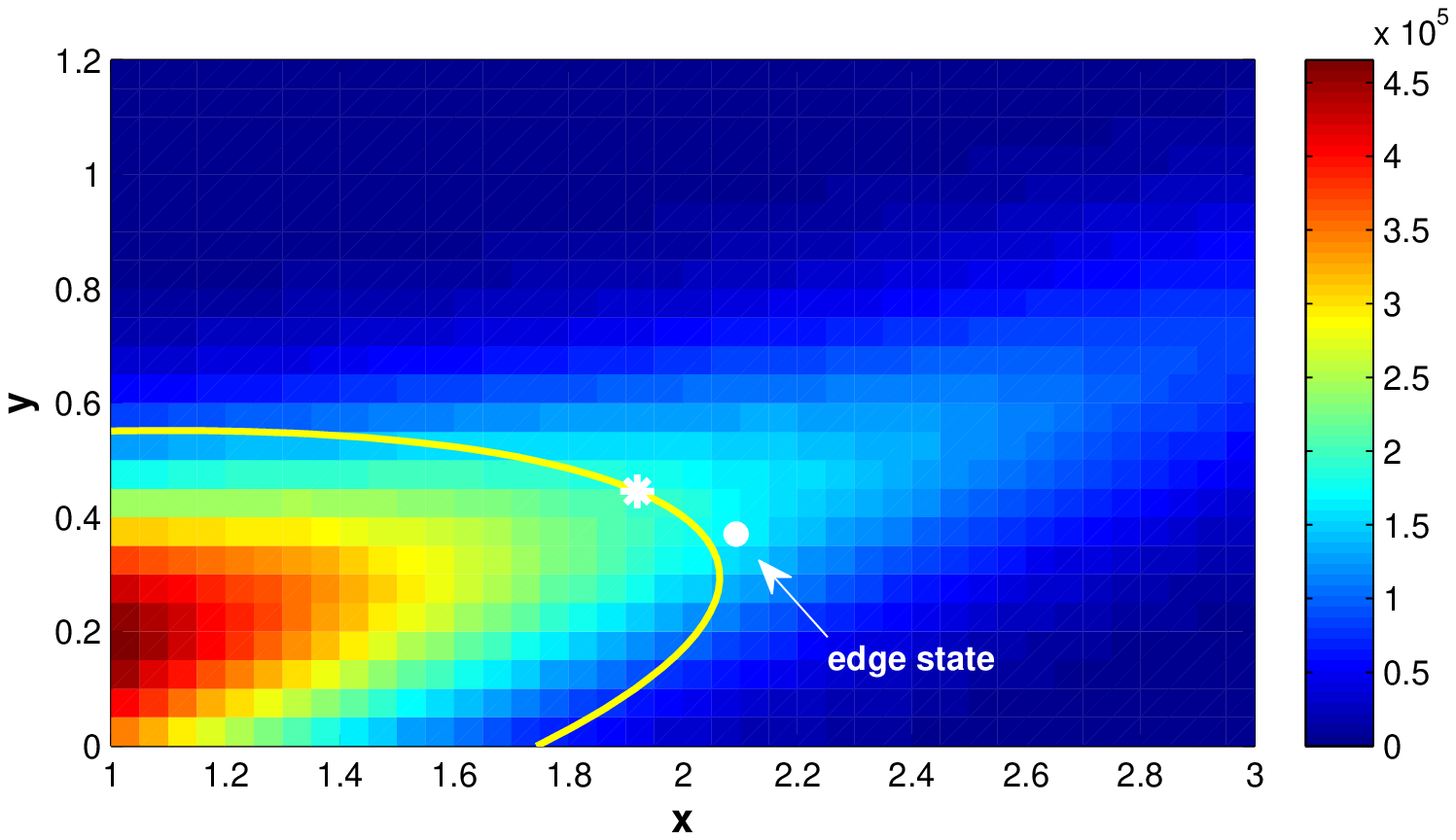}
\caption{\label{fig5} 
Probability density as in Figure \ref{fig4}, but for the full nonlinear equations with noise. Note that the iso-contours
of equal probability are stretched out towards the turbulent state and that they cross the stable manifold
close to the point of contact indicated as the optimal state.
}
\end{figure}

In the noisy case, transition is induced when a fluctuation carries the system across the 
stable manifold. A good estimate of the likelihood of transition can be obtained by considering the probability density at the transition point. Given the functional form of the pdf, the biggest contribution to its variations comes from the quadratic form in the exponent. The equation shows that the iso-contours $p=const$ are ellipsoids determined by $Q_N=const$ that decrease or increase with the noise amplitude $D$. Therefore, if we want to describe where a noisy trajectory crosses over to the turbulent state,
we again have to study iso-contours of a quadratic form, $Q_N=const$, and determine where they touch
the stable manifold of the saddle state. In contrast to the energy functional \eqref{eq:energy} and the dissipation functional \eqref{eq:dissip}, the fluctuation functional $Q_N$ depends on the Reynolds number.
The point of contact between the probability iso-contours and the stable manifold then corresponds
to the point where trajectories are most likely to cross over the stable manifold and to become turbulent.
Alternatively, if one wants to push the system to become turbulent, small perturbations in that
region are most effective because the border is so close.

Figures \ref{fig4} and \ref{fig5} show the relative probability density to be at $(x,y)$ in the region where the transition is expected to occur. It is obtained by integrating $20000$ initial conditions in time for 20 time units starting at the laminar state with a step size of $dt=10^{-3}$. As the phase space is symmetric with respect to the origin, trajectories from the third quadrant are mirrored into the first quadrant. Figure \ref{fig4}, obtained without the nonlinear part, shows the Gaussian shape of the iso-contours. Out of the $400 \times 10^6$ calculated points of the trajectories, more than $108 \times 10^6$ lay in the plotted region of phase space. In Figure \ref{fig5} the nonlinear part is added and the pdf stretches out along the path to the turbulent state.
The figure shows clearly that this happens close to the point where the ellipsoid $Q_N(x,y)=const$ touches the stable manifold. Here more than $75 \times 10^6$ points lay in the interesting region of phase space.
We note that the shape of the iso-contours of the pdf is independent of the noise amplitude $D$ (within the linear approximation), so that changes in $D$ will predominantly influence the likelihood of a transition, but not the path it takes. 

\begin{figure}
\includegraphics[width=\columnwidth]{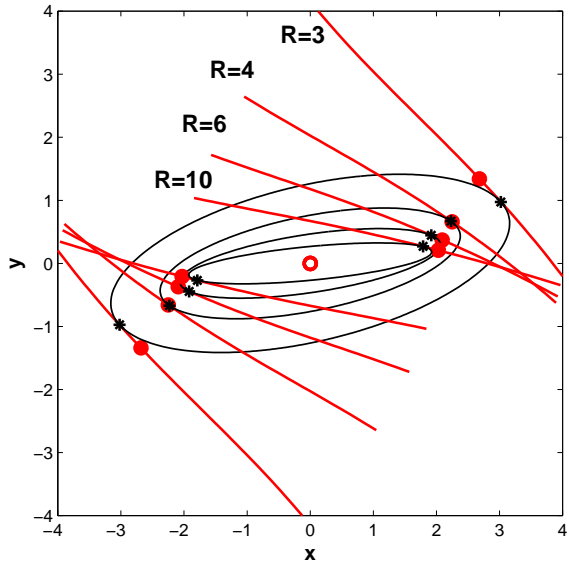}
\caption{\label{fig6} 
Isocontours of the probability density function for different $R$. Note that
as $R$ increases, the ellipsoids of the iso-contour become narrower in the
$y$-direction and stretch out along the $x$-direction. In combination with
the rotation of the stable manifolds (red lines) the point of contact (stars) now stays close to the 
edge state (full symbols) and moves towards the $x$-axis. This is physically plausible,
as a small perturbation in $y$ (in the vortex direction) will produce a strong
streak in the $x$-direction, and it is then the streak that triggers the transition.
}
\end{figure}

More examples of such iso-contours are shown in Figures \ref{fig1} and \ref{fig6} for different values of $R$. With increasing $R$ the ellipsoids become more elongated in $x$-direction, as a result of the asymptotic behaviour of the quadratic form
noticed above (\ref{QN_def}). As they are rotated in the direction opposite to the rotation of the stable manifolds, the point
of contact stays close to the edge state. Within the hydrodynamic interpretation, the transition is dominated
by the streaks ($x$-component) that form as a result of the vortices ($y$-component), 
not by the vortices themselves. 

\section{Summary and conclusions}
The calculations illustrate how different optimization criteria select different optimal initial conditions
for the transition to turbulence. Geometrically, this is to be expected since
different quadratic forms give rise to different ellipsoids in their iso-contours and hence
also different points of contact with the stable manifolds. We note that the results of \cite{Rabin:2012ib}
suggest that for optimization with a time-integrated functional the difference between
energy and dissipation functionals are smaller and may actually vanish. However, we have not 
pursued this question further.

\begin{figure}
\includegraphics[width=\columnwidth]{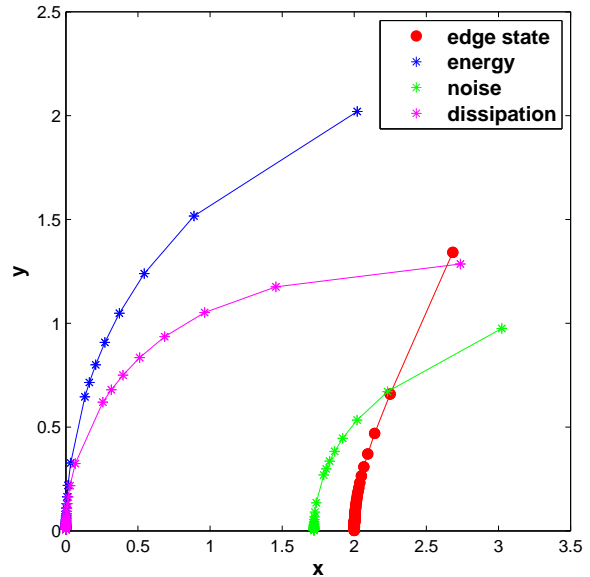}
\caption{\label{fig7} 
Location of the points of contact. The top graph shows
that for the energy and dissipation functional the optimal perturbations are 
vortex like and move towards the origin along the $y$-axis. The noise-optimals
are more streak-like and move towards the $x$-axis, and actually remain close
to the edge state. The edge state (red bullets) approaches $(2,0)$ as $R\rightarrow\infty$.}
\end{figure}

The variation of the optimal points of contact is summarized in Fig. \ref{fig7}.
The data indicates that while the optimal perturbations
are vortex like for the energy and the dissipation functional, they are streak like for the
noisy transition. The difference can be rationalized by the different dynamics. In the deterministic cases,
with the energy and the dissipation functional, small vortex like initial conditions can grow in time to develop
the streaks which then drive the transition. The noisy system is always exposed to small perturbations
which can grow to develop streaks, so that the pdf is elongated in the streak direction by non-normal
amplification. Therefore, the transition happens on top of the already existing streaks and
noise driven flows 
\cite{Sengers:2010jz,OrtizdeZarate:2011kk,OrtizdeZarate:2012dq}
may show different structures at the point of transition than flows driven by judiciously chosen 
initial conditions.  

\begin{figure}
\includegraphics[width=\columnwidth]{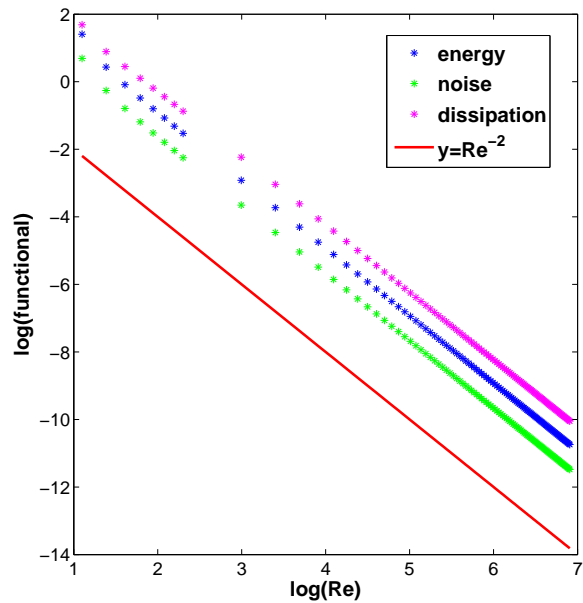}
\caption{\label{fig8} 
Scaling of the functionals for the optimal perturbations. Even though the point of contact moves
differently, the critical values decay like $1/R^2$ in all three cases.
}
\end{figure}

A final quantity to study is the scaling of the functional with Reynolds number, as shown in Fig. \ref{fig8}.
Despite the differences in dynamics, the functionals scale in all three cases like $1/R^2$ for 
large $R$. The particular exponent is specific to the model studied here and the 
type and form of the nonlinear interactions, as other nonlinearities can require
a rescaling near the origin \cite{Eckhardt:2006}. However, the fact that all three cases show the
same scaling could also apply to the full flow cases, as it is a consequence of the measure
used and not the particular nonlinearity at play. What the model also shows is that
deviations from the asymptotic behavior appear close to the point of bifurcation. It is
tempting to speculate that such effects may be responsible for the different
critical exponents that have been observed in pipe flow or plane Couette flow, but that
clearly requires the transfer of the present analysis to realistic flow simulations
and a careful analysis of the asymptotic properties. 

The analysis of simple models has repeatedly helped to elucidate many features of the 
transition to turbulence in shear flows, and to develop tools to explore them
\cite{Gebhardt:1994,Waleffe:1995,DauchotManneville97,Eckhardt:1999,Moehlis:2004il,Moehlis:2005gy,Lebovitz:2009fb,Lebovitz:2013,Kerswell:2014vk}. It is in this spirit that we have
used a forward integration technique to find the optimal points on the stable manifolds 
for different functionals and to explore the changes with Reynolds number. We expect that 
many of the features described here can also be found in the high-dimensional state spaces
of realistic shear flows, perhaps after suitable modifications and adaptations of the methods
used to explore the high-dimensional spaces.

\section*{Appendix}
In this appendix we discuss the modification of the edge tracking algorithm \cite{Skufca:2006iu} 
used for the determination of the initial conditions on the edge that optimize a prescribed 
quadratic functional $Q_N(x,y)$.
The functional may be the energy \eqref{eq:energy}, the dissipation \eqref{eq:dissip} 
or the argument in the pdf \eqref{eq:Qnoise}.
To keep the notation compact, we denote the equations of motion in vectorial notation
as $\dot {\bf x}={\bf f}$.

We begin with an arbitrary initial condition in the vicinity of the edge and we let it evolve in time towards the edge state. Unlike other edge tracking methods, where trajectories are integrated until they are sufficiently close to the laminar or the turbulent state, we here stop the integration at the time when the distance to the edge state is minimal. The trajectory's velocity at the turning point is then projected onto the normal of the stable eigenvector to decide if the tested initial condition moves upwards
or downwards, towards the turbulent or the laminar state. With this criterion we can determine a point ${\bf x}_0$ on the edge
also when the point is very close to the edge and the time needed to pass the edge state becomes excessively large.

We then propagate this point along the time direction, 
${\bf x}_1= {\bf x}_0 + s {\bf f}({\bf x}_0)$ by an amount $s$ that is chosen such that
$Q(s)$ is minimized. Formally,
\begin{equation}
s=-\frac{{\bf f}^T({\bf x}_0) \, Q \, {\bf x}_0 + {\bf x}_0^T \, Q \, {\bf f}({\bf x}_0)}{2 {\bf f}^T({\bf x}_0) \, Q \,  {\bf f}({\bf x}_0)}.
\end{equation}
In numerical implementations, $\|s{\bf f}\|$ is kept below a certain threshold to stay in a
region where linear approximations are possible. Then a new edge tracking is
started from ${\bf x}_1$ and the process is repeated until the norm of the total shift $\|s{\bf f}\|$ falls below
a convergence threshold, here $10^{-5}$.

{\small \textit{This work was supported in part by the German Research Foundation.}}


\end{document}